\documentstyle[12pt]{article}
\def\by#1#2{{\displaystyle {#1}\over \displaystyle {#2}}}
\def\d{{\rm d}}
\newcommand{\al}{\alpha}
\newcommand{\alv}{\alpha_V}
\newcommand{\als}{\alpha_S}
\newcommand{\be}{\beta}
\newcommand{\bev}{\beta_V}
\newcommand{\bes}{\beta_S}
\newcommand{\ga}{\gamma}
\newcommand{\gav}{\gamma_V}
\newcommand{\gas}{\gamma_S}
\newcommand{\alb}{\overline{\alpha}}
\newcommand{\beb}{\overline{\beta}}
\newcommand{\gab}{\overline{\gamma}}
\begin{document}

\begin{flushright}
MRI-PHY/P980238 \\
\today \\ [0.8cm]
\end{flushright}

\begin{center}
{\Large \bf An SU(3) model for octet baryon and meson fragmentation} \\
[0.2cm]
D. Indumathi${}^1$, H.S. Mani${}^1$, Anubha Rastogi${}^{1,2}$ \\ [0.5cm]
${}^1${\it Mehta Research Institute of Mathematics and Mathematical
Physics, Jhusi, Allahabad 211 019, India} \\
${}^2${\it Dept. of Physics and Astrophysics, University of Delhi,
Delhi 110 007} \\ [2cm]

{\bf Abstract} \\ [0.3cm]
\end{center}

The production of the octet of baryons and mesons in $e^+\,e^-$
collisions is analysed, based on considerations of SU(3) symmetry and a 
simple model for SU(3) symmetry breaking in fragmentation functions. 
All fragmentation functions, $D_q^h(x, Q^2)$, describing the fragmentation
of quarks into a member of the baryon octet (and similarly for
fragmentation into members of the meson octet) are expressed in terms of
three SU(3) symmetric functions, $\alpha(x, Q^2)$, $\beta(x, Q^2)$, and
$\gamma(x, Q^2)$. With the introduction of an SU(3) breaking parameter,
$\lambda$, the model is successful in describing hadroproduction data
at the $Z$ pole. The fragmentation functions are
then evolved using leading order evolution equations and good fits to
currently available data at 34 GeV and at 161 GeV are obtained.

\vspace{0.5cm}


\vspace{1.0cm}

\section {Introduction}

The formation of hadrons from fragmentation of partons is of
considerable current interest \cite{Vogelsang,Kramer}.
While the production of partons in any
process---be it Deep Inelastic Scattering or $e^+e^-$
annihilation---can be calculated, perturbative QCD can only predict
the scale ($Q^2$) dependence of the process of fragmentation of quarks
and gluons into hadrons; the fragmentation functions 
themselves are not perturbatively calculable and can only be modelled. 
Various models exist which attempt to explain the process of fragmentation
\cite{Lund,Nzar,Chun}. Many computer simulations \cite{Jetset,Herwig}
also exist and are in popular use. However, the r\^ole of strangeness
suppression as well as isospin in hadroproduction is not yet clearly
established. A clean channel to study such phenomena is provided by
$e^+e^-$ annihilation experiments due to the fact that the initial
interacting vertex is purely electromagnetic in nature. These
experiments have been performed at different energies
\cite{Data90,Sigdata,Data34}.

We propose a simple model for a light quark ($u$, $d$, $s$) to fragment
into an octet
baryon or a pseudoscalar meson, using SU(3) symmetry of quarks and octet
hadrons. All fragmentation functions are described in terms of three 
SU(3) symmetric functions $\al(x,Q^2)$, $\be(x,Q^2)$ and $\ga(x,Q^2)$
(one set for baryons and another set for mesons) and an
SU(3) breaking parameter $\lambda$ which have been determined by comparison 
with data. The model is able to predict the $x$-dependence of all octet 
baryons and mesons. There is good agreement with data 
at two different energies (corresponding to $Z^0$ and photon exchange)
over most of the $x$ range of available data. Hence, the overall success
of the
model does seem to indicate the existence of an underlying SU(3) symmetry
between members of a hadron octet.

The paper is organised as follows: in the next section, we develop the
model for quark fragmentation into octet baryons and mesons. In Section
3, we fix our model parameters using data on some hadrons at the $Z^0$
pole and
use the resulting fits to predict production rates for other hadrons. We
find good agreement with data. In Section 4, therefore, we use leading
order evolution equations to simultaneously fit data at two different
energies corresponding to hadroproduction via $Z^0$ and photon exchange. 
We use both the standard DGLAP evolution equations \cite{Dglap} as well 
as the modified leading log approximation (MLLA) \cite{Dok} to evolve 
the fragmentation functions
to different energies and compare the model with data. Section 5
contains discussions on our results and concludes the paper. 
Some details for the expressions of the different quark fragmentation
functions in terms of $\al$, $\be$ and $\ga$ are given in Appendix A.

\section{Cross Section and Kinematics}

We consider the production of hadrons in $e^+\,e^-$ annihilations via
$\gamma$ and $Z$ exchange. To leading order, the cross-section for
producing a hadron $h$ can be expressed \cite{Nasweb} in terms of the unknown
fragmentation functions, $D_q^h(x_E, Q)$, as
\begin{eqnarray}
\by{1}{\sigma_{tot}}\by{\d \sigma^h}{\d x_E} & = & \by{\sum_q c_q
D_q^h(x_E, Q)}{\sum_q c_q}~.
\label{eq:totcross}
\end{eqnarray}
Here $c_q$ are the charge factors associated with a quark $q_i$ of
flavour $i$ and can 
be expressed \cite{Nasweb} in terms of the electromagnetic charge,
$e_i$, and the vector and axial vector electroweak couplings, 
$v_i = T_{3i}-2 e_i \sin^2 \theta_{\rm w}$ and $a_i = T_{3i}$ as
\begin{eqnarray} \nonumber
c_q & = & c_q^V + c_q^A~, \\ \nonumber
c_q^V & = & \by{4 \pi \alpha^2}{s}[e_q^2 + 2e_q v_e v_q \, \rho_1(s)
+(v_e^2 + a_e^2)v_q^2 \, \rho_2(s)]~, \\ 
c_q^A & = & \by{4 \pi \alpha^2}{s}(v_e^2+a_e^2)a_q^2 \rho_2(s)~, \\ \nonumber
\rho_1(s) & = & \by{1}{4 \sin^2 \theta_{\rm w} \cos^2 \theta_{\rm w}}
\by{s(m_Z^2-s)}{(m_Z^2-s)^2+m_Z^2 \Gamma_Z^2}~, \\ \nonumber 
\rho_2(s) & = & \left (\by{1}{4 \sin^2 \theta_{\rm w} \cos^2 \theta_{\rm
w}}\right )^2 \by{s^2}{(m_Z^2-s)^2+m_Z^2 \Gamma_Z^2}~. \nonumber
\end{eqnarray}
In Eq. \ref{eq:totcross}, a sum over quarks as well as
anti-quarks is implied. Here $x_E$ is the energy fraction, $x_E =
E_{\rm hadron}/E_{\rm beam} = 2 E_h/\sqrt{s}$. We shall also use the
momentum variable,
$x_p = P_{\rm hadron}/P_{\rm beam} = 2 P_h/\sqrt{s}$; $ x_E^2 = x_p^2
+ 4m_h^2/s$,
where $m_h$ is the mass of the hadron $h$ and $Q$ is the energy scale of
the interaction and is equal to $\sqrt {s}$. We shall normally use $x$
to mean $x_E$ unless otherwise specified. 

The fragmentation function, $D_q^h(x_E,Q)$, associated with the quark
$q$ is the probability for a quark $q$ to hadronise to a hadron $h$
carrying a fraction $x_E$ of the energy of the fragmenting quark.
This is not perturbatively calculable from theory, although the scale
($Q$) dependence of these functions is given by QCD. Data from different
experiments at different energies from $\sqrt{s} = 10$--$91.2$ GeV exists
on  $p$, $\Lambda$, $\Sigma$ and $\Xi$ octet baryon production as
well as on $\pi$, $K$, $\eta$ octet meson production
\cite{Data90,Sigdata,Data34}. 
All available data measures the production rate of hadron plus antihadron. 
Due to the symmetric nature of the process $e^+\,e^- \to q \overline q$,
the resulting hadron and antihadron yields are equal. We therefore
present results for the sum of hadron and antihadron yields in what
follows.

We can also re-express the cross section in terms of the octet and
singlet fragmentation function combinations, as
\begin{eqnarray}
\by{1}{\sigma_{tot}} \by{\d \sigma^h}{\d x_E} & = & \by{a_0
\Sigma^h(x_E,t) +a_3 D_3(x_E,t) + a_8 D_8(x_E,t)}{\sum_q {c_q}} ~,
\label{eq:sns}
\end{eqnarray}
where $\Sigma$, $D_3$ and $D_8$ refer to the singlet, and the two octet
($(u - d)$ and $(u + d - 2 s)$) combinations respectively, with $a_0 =
(c_u+c_d+c_s)/3$; $a_3 = (c_u-c_d)/2$ and $a_8 = (c_u+c_d-2c_s)/6$. 
 
We now present our model for quark fragmentation functions. 

\section{The Model}

We study semi-inclusive production of the octet baryons  and
the pseudo-scalar octet mesons in $e^+\,e^-$ annihilation process 
using SU(3) symmetry of the quarks and of the hadrons in their respective 
octets. The production of the entire meson and baryon octet is described
in terms of SU(3) symmetric quantities. We study only light quark ($u$,
$d$, $s$) fragmentation where the fragmenting quark ($q_i$) is a member
of the quark triplet ($q_1 = u$, $q_2 = d$, $q_3 = s$) and the
hadron under study ($h_i^j$) is a member of the baryon (or meson) octet
(see Table \ref{tab:octet}):
$$
q \rightarrow h + X~,
$$
with $X$ being a triplet, antisixplet or fifteenplet. Since the final
states in these three processes are distinct, we can express all the
quark fragmentation functions for the hadron $h$ in terms of the {\it
independent} fragmentation probabilities $\alpha(x, Q)$, $\beta(x, Q)$,
and $\gamma(x, Q)$ where $\alpha$ ($\beta$, $\gamma$) is the fragmentation
probability when $X = 3 (\overline{6}, 15)$. The corresponding
probabilities for antiquark fragmentation are $\overline{\alpha}$,
$\overline{\beta}$ and $\overline{\gamma}$ (in this case, there is an 
antitriplet fragmenting into an octet and $X$, with the $X$ being a 
$\overline{3}$,6 or $\overline{15}$). These probabilities are also
functions of $x_E$. The quark and antiquark fragmentation into $h_i^j$,
$i,j = 1,\ldots, 8$, in terms of the SU(3) probability functions, $\alpha$,
$\beta$, and $\gamma$ are given in Table \ref{tab:frag}. Note that
the functions $\alpha$, $\beta$ and $\gamma$ for the baryon and meson
octets are unrelated; they just correspond to the same underlying symmetry. 

Since the (more) massive strange quark is known to break SU(3) symmetry,
we introduce symmetry breaking effects as follows: the fragmentation
function is suppressed by an $x_E$--independent factor $\lambda$
whenever a strange quark
belonging to the valence of the hadron is produced. This means
that all non-strange fragmentation functions of strange hadrons are
suppressed by $\lambda$. For example, $D_u^K$ or $D_u^\Lambda$ are
suppressed by a factor $\lambda$ compared to $D_s^K$ or $D_s^\Lambda$.
Note that all (strange and nonstrange) sea fragmentation functions
corresponding to a given hadron come with the {\it same} factor of
$\lambda$.

There are 3 quark and 3 antiquark fragmentation functions
for each hadron.  Since there are eight baryons (or mesons) in the octet,
this corresponds to a total of forty eight unknown functions for
a given octet, which, in
principle, need to be fitted to data. However, in our model, all the
hadron fragmentations for a given octet can be described in terms of
6 functions
alone, along with an SU(3) breaking parameter, $\lambda$, thus leading
to an enormous simplification in the analysis as well as dramatically
increasing the predictive power of the model.

We now go on to detail the model, first for the case of octet baryons
and then for the octet mesons.

\section{Comparison with data at the ${\bf Z}$ pole}

We relate the model parameters and try to find $\alpha$, $\beta$,
$\gamma$ (and the corresponding antiquark functions) as well as the SU(3)
breaking parameter $\lambda$ by comparing with data on $e^+\;e^- \to Z^0
\to q \overline q$ obtained at LEP \cite{Data90,Sigdata}.

\subsection{Baryon Fragmentation}

First of all, we observe that our model predicts equal rates of production
of $\Sigma^0$ and ($\Sigma^+ + \Sigma^-$)/2 (See Table \ref{tab:frag}) due
to isospin symmetry. This is borne out by data \cite{Sigdata}; for
instance, the multiplicities of ($\Sigma^+ + \Sigma^-$)/2 and $\Sigma^0$
at $Q = 90$ GeV are $0.091 \pm 0.019$ and $0.071 \pm 0.018$ respectively
and are compatible\footnote{Our model does not account for isospin
breaking effects which are small compared to SU(3) breaking effects but
are known to exist; hence we are looking for agreement only to within
this approximation.} within $1\sigma$. 

To obtain predictions for the other baryons, we separate the quark
fragmentation functions into valence and sea parts by defining, as usual,
\begin{eqnarray} \nonumber
\al = \alv+\als~; & \be  = \bev+\bes~; & \ga  = \gav+\gas~; \\
\alb  = \als~; & \beb  = \bes~; & \gab  = \gas~.
\label{eq:valsea}
\end{eqnarray}
There is no $s$ quark in the valence of the proton; hence, we obtain 
$\gav$ = 0 or $\ga$ = $\gab$ = $\gas$ (see Table \ref{tab:frag}).
Furthermore, using $D_{u_s}^\Lambda = D_{d_s}^\Lambda = D_{s_s}^\Lambda$,
leads to the constraint, $\beb = \ga/4 + \alb/3$~. The final simplifying
assumption of $D_{\overline u} =  D_{\overline d}$ for all baryons
allows us to express all antiquark fragmentation functions in terms of
$\gamma$ alone:
\begin{eqnarray} \nonumber
\gab & = & \ga~; \\ 
\alb & = & 0.75 \ga~; \\ \nonumber
\beb & = & 0.5 \ga~; \nonumber
\end{eqnarray}
The model for octet baryons therefore has three unknown functions
$\alv(x_E, Q^2)$, $\bev(x_E, Q^2)$, $\ga(x_E, Q^2)$ and an unknown
parameter, $\lambda$, that characterises SU(3) symmetry breaking.
We now attempt to evaluate these by suitable comparison with data.

\begin{figure}[ht]
\centering
\vskip 17truecm

{\includegraphics{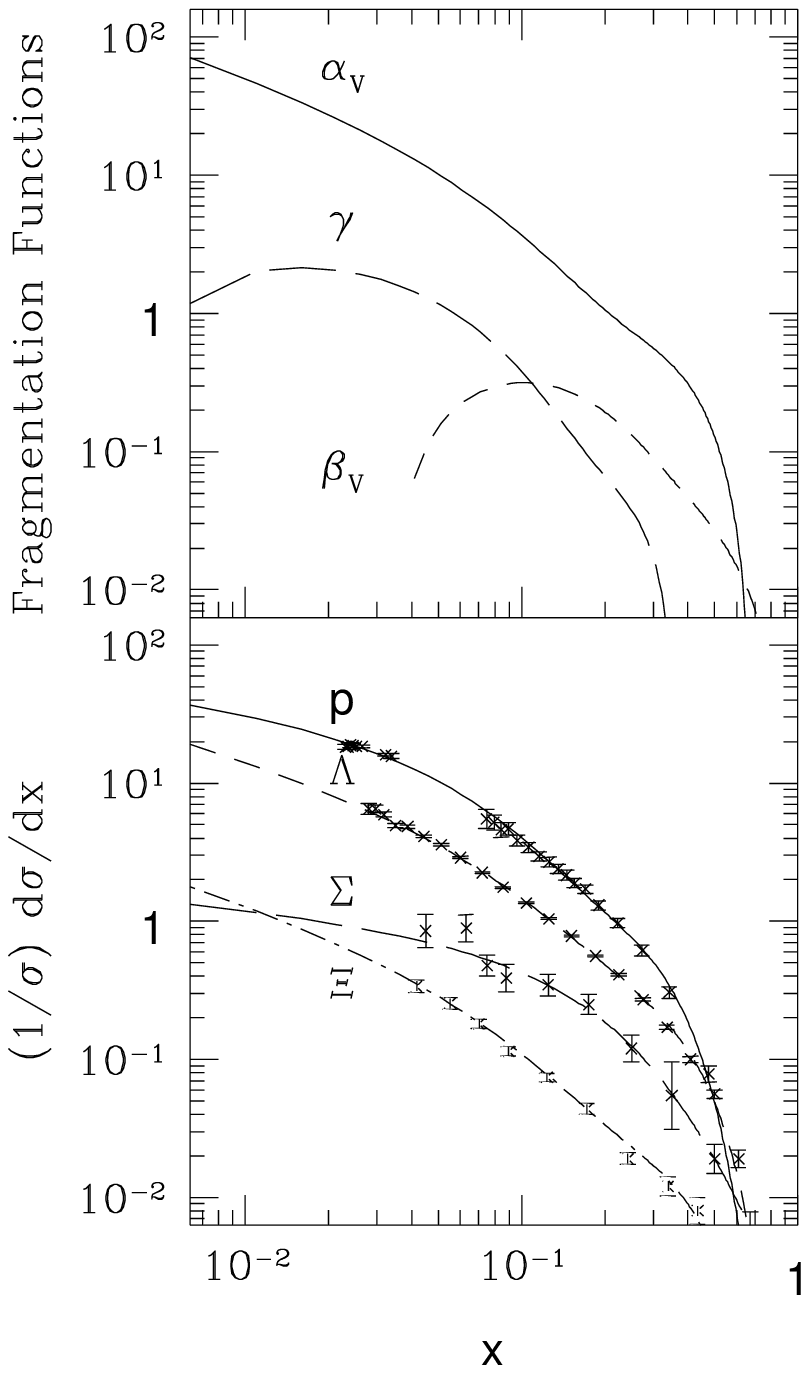}}

\includegraphics{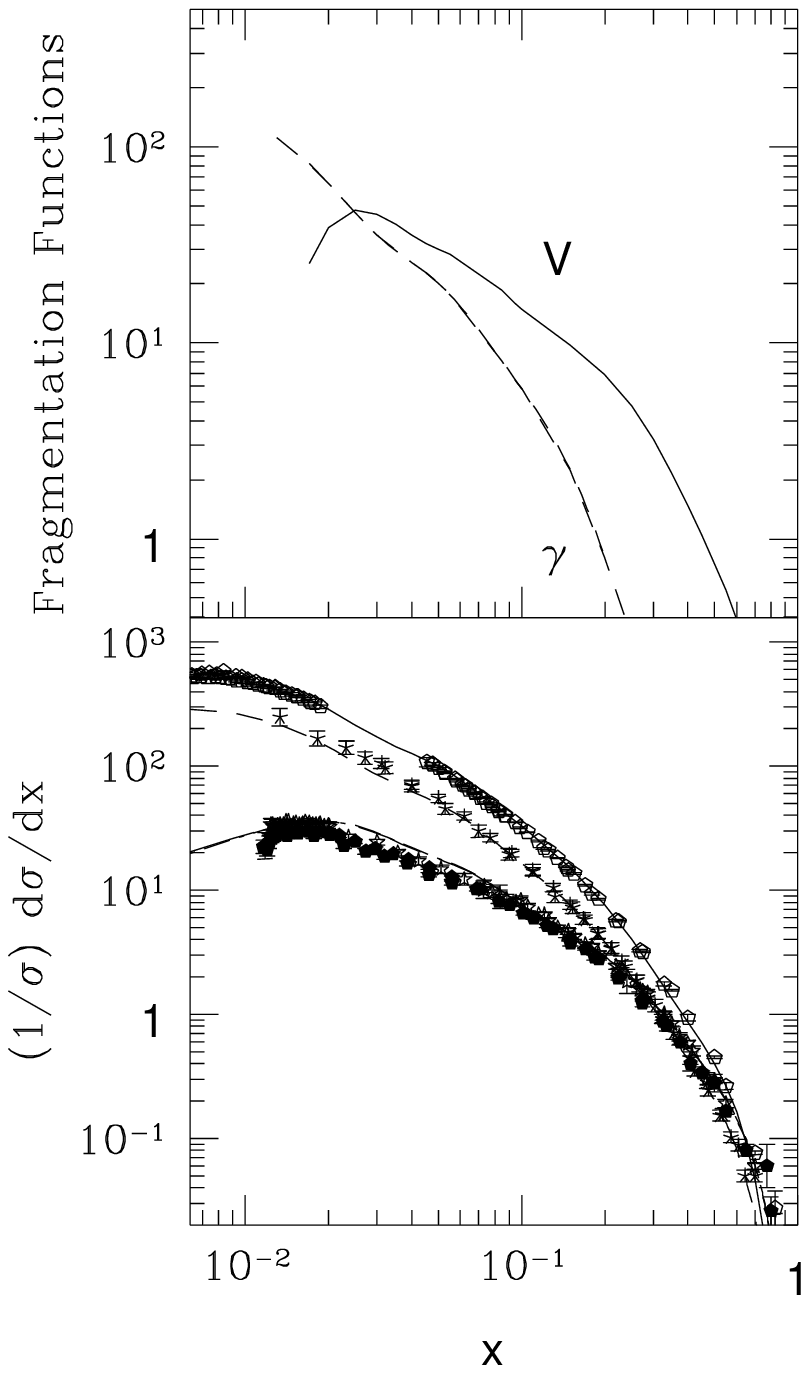}
\caption[dummy]{The figure on the left shows the baryon
fragmentation probabilities $\alv$, $\bev$
and $\ga$, fitted using data on $p$, $\Lambda$ and $\Sigma$ at 90 GeV
\cite{Data90,Sigdata} as a function of $x$, and the prediction for the
$\Xi$ baryon using these, along with the data used. The figure on the
right shows the mesonic probabilities, $V$ and $\ga$,
and the predictions for mesons; the fits shown are to data corresponding
to $\pi^\pm$, $\pi^0$, $K^\pm$ and $K^0$ in decreasing order of
magnitude.}

\label{fig:model}
\end{figure}

We expect valence fragmentation functions (leading quark fragmentation) to
dominate in the large $x_E$ region and sea fragmentation
functions to dominate at small $x_E$. We use the large $x_E$ data to fix
the value of $\lambda$. Since $\Xi^-$ has two $s$ quarks in its valence,
its $s$-valence fragmentation function is suppresesed by a factor of
$\lambda$ compared to $p$, $\Lambda$ and $\Sigma^{\pm,0}$. Therefore, 
we expect $\Xi^-$ production cross sections to be smaller than for
other baryons even in the large $x_E$ range, where only the valence
contribution survives. Indeed, as can be seen from Fig. \ref{fig:model},
data \cite{Data90}
show that the large $x_E$ cross sections are similar for all octet baryons,
$p$, $\Lambda^0$, $\Sigma^{\pm, 0}$, within errorbars, while $\Xi^-$
data is smaller in that region. Furthermore, assuming that $u$ quark
fragmentation dominates the large $x_E$ proton data, we see from Table
\ref{tab:frag}
that the ratio of $\Xi^-$ to proton production rates is just $\lambda$
(the expressions for the two are otherwise the same). Using the data at
$\sqrt{s} = 91.2$ GeV, we find that $\lambda = 0.07$. This result can be
corrected due to the fact that $D_d^p$ is not small at large $x_E$;
however, we shall see that this value for $\lambda$ gives good agreement
with data. 

Since $\lambda$ turns out to be quite small, the production rate of
strange baryons is dominated by $s$-quark fragmentation. This means that
the $\Lambda^0$ rate is sensitive to $\alpha_V$ while the $\Sigma$
baryons give information on $\beta_V$ (see Table \ref{tab:frag}). On
the other hand,
$p$ and $\Xi$ depend on the combination $(\alpha_V + \beta_V)$; it is
therefore possible to separately determine $\alpha_V$ and $\beta_V$
from the data, rather accurately, especially in the valence-dominated
region of large $x, x$ \raisebox{-0.4ex}{$\stackrel{>}{\scriptsize \sim}$}
$0.1$.
The (SU(3) symmetric) sea is the same for all members of the octet, upto
overall powers of lambda. 

Note that the measured hadron spectrum is an inclusive one. We take this
into account, especially for the case of $p$ and $\Lambda$, by defining
inclusive fragmentation functions in terms of the exclusive ones used so
far:
\begin{eqnarray*}
p_{\rm inclusive} & = & p_{\rm exclusive} + 0.52 \Sigma^+ + 0.64 \Lambda~; \\
\Lambda_{\rm inclusive} & = & \Lambda_{\rm exclusive} + 1.0 \Sigma^0 +
                        1.0 \Xi^- + 1.0 \Xi^0~.
\end{eqnarray*}
The multiplying fractions indicate the branching fractions into $p$ and
$\Lambda$ of the various baryons. 
Here $\Sigma^*$ and $\Lambda_c$ decays to the various baryons have been
ignored since they are very small. $\Lambda_c$ data is between 15 to 100
times smaller than $\Lambda$ and proton data in the overlapping $x_E$ range
of about 0.3--0.8 \cite{Lamcmult}. The $\Sigma^{\pm}$ and $\Xi^-$ data is 
considered to be purely exclusive for this reason. The energy ($x_E$) of the
daughter-baryon is taken to be the same as that of the parent
because of the small difference in masses of $p$, $\Lambda$, $\Sigma$ and
$\Xi$. 

For the case of the 91.2 GeV $Z$ exchange data, $\alv$, $\bev$ and $\ga$
were evaluated at different $x_E$ values using parametrisations to
the $p$, $\Lambda$ and $\Sigma$ data \cite{Data90,Sigdata} and the value of
$\lambda$ as already estimated; these were then used to predict the
cross section for $\Xi^-$. We find that the data is fitted very well at
all $x_E$ as is shown in Fig. \ref{fig:model}, where the
resulting fits to $\alv$, $\bev$ and $\gamma$ are also shown. Note that
a change in $\lambda$ can alter the overall normalisation but not the shape
of the distribution, which is a prediction of this model. 

\subsection{Meson Fragmentation}

The pseudo-scalar meson-octet is a self conjugate octet i.e., the
same octet contains mesons as well as their antiparticles. The mesons
and their antiparticles are related by charge conjugation. This means
that $D_q^h = D_{\overline{q}}^{\overline{h}}$. As a consequence of this
fact we immediately see that $\alb$, $\beb$ and $\gab$ are not independent
of $\alpha$, $\beta$ and $\gamma$ (see Table \ref{tab:frag}). Of these
six quantities, only three are independent. We choose these to be
$\alpha$, $\beta$ and $\gamma$.
Due to mixing with the singlet sector, we do not consider
the $\eta$ meson here. 

Just as in the case of $\Sigma$, here we predict equal rates for $(\pi^+
+ \pi^-)/2$  and $\pi^0$, due to isospin invariance. This is also borne
out by data \cite{Data90}. We therefore consider only the combinations
$m^{\pm} \equiv (m^+ + m^-)$ for $m = \pi, K$, and $(K^0 +
\overline{K^0}) \equiv 2 K_s^0$. 

We make the usual separation into valence and sea fragmentation
functions. As before, we reduce the number of unknown functions
through various symmetry considerations. We assume that the sea is SU(3)
flavour symmetric, so that $D_u^{\pi^-}$
= $D_s^{\pi^-}$ and so on. Using this, we have $\beta$ = $\gamma$/2 and
all sea fragmentation functions are equal to $S = 2\gamma$. 

Thus, all valence fragmentation functions can be expressed in terms of
the function, $V$, where $V$ is given, for example, by the difference
$(D_u-D_{\overline u})$ in $\pi^+$, as
$$
\begin{array}{lclcl}
V & = & \alpha + \beta - {5 \over 4}\gamma & = & \alpha - {3 \over 4}
\gamma ~,
\end{array}
$$ 
and all sea fragmentation functions in terms of $\gamma$. Now $V$ and
$\gamma$ can be determined by comparison with ($\pi^{\pm}$, $K^{\pm}$ and
$K^{0}$) data; since the two sets of $K$ data are not very different
\cite{Data90}, it is not possible to determine $\alpha$ and $\beta$
individually in this case.

The assumption (made in the baryon case) that the daughter hadron carries 
away the bulk of the energy of the parent, thus contributing to the same
$x_E$ bin, enabled us to simply add the various hadron fragmentation
functions to arrive at an ``inclusive'' hadron fragmentation function.
Here, since both $\pi$'s and $K$'s are very much lighter than their
decay sources (mostly $D$ mesons and baryons), this assumption is no
longer reasonable. We therefore merely estimate errors
arising from the inclusive nature of the data by comparing multiplicities
rather than $x_E$ distributions. 

The bulk of the contamination of the pion sample which is due to $K_s^0 \to
\pi\, \pi$ decays is estimated to be about 4\% from the
ratio of the relative multiplicities \cite{Data90,Mult}, $N^{\pi}/N^{K_s}
\sim 10$ and a 
branching fraction $\sim 0.35$ for the decay for $\pi = \pi^+, \pi^-,
\pi^0$; hence we ignore this. $K_L^0$ does not decay
within the detector. However, there is a substantial contribution from
charm meson feeddown for the $K$ data sample (from the decay of all the
$D$s); this is estimated (from multiplicity data \cite{Data90,Mult}) 
to be about 16\% for $K^\pm$ and about 20\% for $K^0$ mesons. 
The contamination from baryons is negligible.

As before, we include SU(3) symmetry breaking effects: the
fragmentation probability is suppressed by a factor $\lambda$ whenever a 
strange quark belonging to the valence of the meson is produced. Since
the fragmenting quark excites a quark pair rather than a diquark pair
(as in the case of baryons) the value of $\lambda$ here is not related
to that for the baryonic sector. We can get bounds for $\lambda$ by
using the $\pi^\pm$ and $K^\pm$ multiplicities $= 17.05 \pm 0.43$ and 
$2.26 \pm 0.18$ respectively \cite{Data90,Mult}: The total rates for
$\pi^\pm$
and $K^\pm$ production (their multiplicities) are related to the first
moment of $V$ and $S$, and the parameter $\lambda$; positivity
constraints on $V$ and $S$ (since they are probabilities) then require
that $\lambda < 0.14$. A tighter constraint on $\lambda$ will be found
in the next section, when we apply the model to data at different
energies. 

We use a typical value of $\lambda = 0.08$ to fit $V$ and $S$ over the
available
$x_E$ range of $\pi^{\pm}$ and $K^{\pm}$ data. These were then used to
predict cross sections for $\pi^0$ and $(K^0 + \overline{K^0})$. The
data is fitted
very well by these functions and is shown in Fig. \ref{fig:model}; the fits to
$V$ and $\ga$ as determined from $\pi^\pm$ and $K^\pm$ are also shown here. 

We have been able to explain the production of the entire meson and baryon 
octet by using SU(3) symmetry and the suppression factor $\lambda$ at 
$\sqrt{s} = 91.2$ GeV. Encouraged by this success, we investigate whether 
the model works for other c.m. energies also. In the next section, we 
discuss the evolution of these fragmentation functions to different energies.

\section{Comparison with data for photon exchange}

\subsection{Leading log evolution}
Over the last decade, several experiments, performed over a wide range of 
c.m. energies, (10 GeV $\leq \sqrt s \leq 91.2$ GeV), have reported 
measurements of cross-section of hadron production. It is known that the
cross-section is a not a constant over the range of c.m. energies; for
instance, this has enabled the extraction of the running coupling constant
$\alpha_s$. We use leading order (LO) DGLAP evolution equations \cite{Dglap}
to relate the fragmentation functions (and hence the cross-section) at a 
given energy $Q$ to those at a different energy $Q_0$. 

Nonsinglet (including the valence functions $\alv$ and $\bev$ for baryons
and $V$ for mesons) and singlet fragmentation functions
evolve differently under evolution. The singlet ($\Sigma^h (x,t) = u(x,t)
+d(x,t)+ s(x,t) +{\bar u}(x,t)+{\bar d}(x,t)+{\bar s}(x,t)$) mixes with
the gluon fragmentation function, $g(x,t)$; here we have used
$x = x_E$, $t = Q^2$, and $D_q^h = q$ for convenience.
Since the gluon is a flavour singlet, we use one gluon fragmentation 
function for all of the mesons and another one for all the baryons.

The symmetry between the singlet sector of different baryons is broken
by $\lambda$. However, all singlet combinations for the other baryons can be 
expressed in terms of $\Sigma^p$ and strange valence fragmentation
functions. We have
\begin{eqnarray} \nonumber
\Sigma^{\Lambda} & = & \lambda \Sigma^p + (1-\lambda)s_V^{\Lambda}~; \\ 
\Sigma^{\Sigma} & = & \lambda \Sigma^p + (1-\lambda)s_V^{\Sigma}~; \\ \nonumber
\Sigma^{\Xi^-} & = & \lambda^2 \Sigma^p + \lambda (1 - \lambda)
s_V^{\Xi^-}~. \nonumber
\end{eqnarray}
Similarly, we construct only the singlet combination $\Sigma^{\pi}$ for the
pion. All other meson singlets can then be recovered from this, and the
valence function $V$. 

The evolution was done in two steps. First of all, for the baryon octet,
the gluon fragmentation function and $\gamma$ (i.e., all the sea quark
fragmentation functions) were set to zero at the starting scale, and
then radiatively generated. This precludes us from having to guess a
possible starting gluon function, which is very poorly (if at all)
constrained. Since the sea of all strange baryons is suppressed by a
factor of $\lambda = 0.07$, only the proton data (with an unsuppressed
sea sector) was poorly fitted.  Fits to other baryons were still fairly
reasonable. This indicates that the fragmentation of, say, $\Lambda$
or $\Sigma$ is dominated by strange quark fragmentation at almost all
$x_E$ values.  Clearly, the function $\gamma$ is necessary to fit the
proton data.  In Ref.~\cite{Nzar}, an analysis for fragmentation functions
for $\Lambda$ and $p$ has been done using SU(6) symmetry of the
baryons. The functions $\widehat{S}$ and $\widehat{T}$ given therein
are analogous to the functions $\alpha$ and $\beta$. However, there is
no function analogous to $\gamma$, which (being the sea contribution)
is required to explain the small $x$ data.

\begin{figure}[ht]
\centering
\vskip 15truecm

\includegraphics{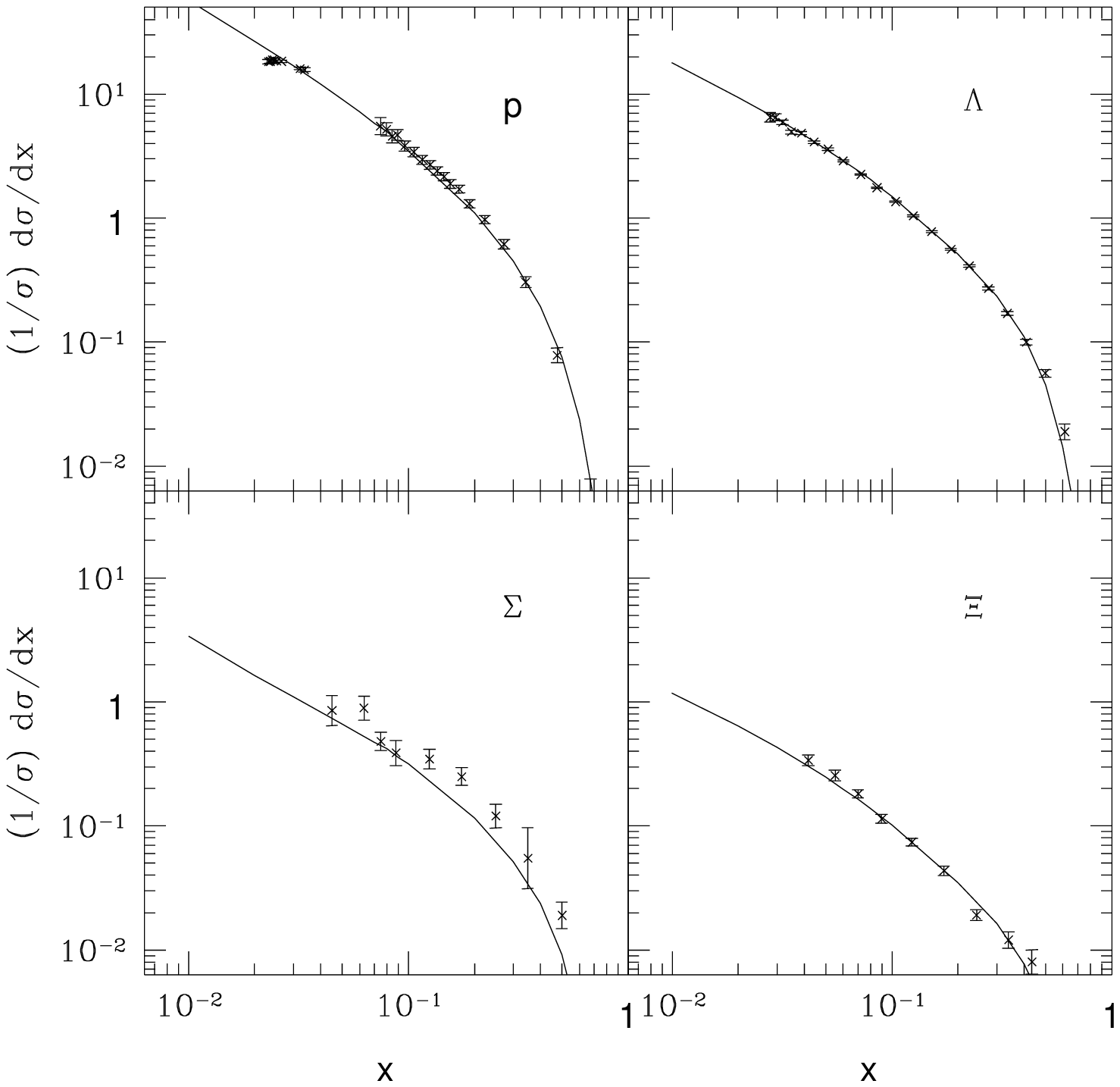}
\caption[dummy]{(a) The figure shows the model fits to the 90 GeV data
for baryons using DGLAP evolution.}

\label{fig:dlla90}
\end{figure}

We then used a small, non-zero starting sea as well as (a
common) gluon fragmentation function to improve the fits, the results of
which are shown in Figs. \ref{fig:dlla90} and \ref{fig:dlla34}. However,
we emphasise that
the evolved fragmentation functions are not very sensitive to the choice
of the gluon fragmentation function, which is therefore not well-determined
in our model.

Each of the fragmentation functions was parametrised at a starting value 
of $Q^2=2$ GeV${}^2$ and evolved up to the final $Q^2 (=s)$ value of the 
data.  The input functions $F_i(x) = \alv, \bev, \gamma$ (for the 
case of baryons) and $F_i(x) = V, \gamma$ (for the case of mesons) were 
parametrised as
\begin{eqnarray}
F_i(x) = a_i(1-x)^{b_i}(x^{c_i})(1+d_ix+e_ix^2)~.
\end{eqnarray}
The parameters $a, b, c, d, e$ for different input fragmentation functions
are given in Table \ref{tab:inputll}.

We tuned the starting parameters to yield a good fit to the 90 GeV
hadroproduction data
which is essentially via $Z$-exchange \cite{Data90,Sigdata}. We then
used the same set to predict the rates for a lower $Q^2$ value which
is dominated by photon exchange.
The resulting fits on the $Z$ pole ($\sqrt{s} = 91.2$ GeV) and a fit to
the available baryon data sample \cite{Data34} at $\sqrt{s} = 34$ GeV for
$\lambda = 0.07$ are shown
in Figs. \ref{fig:dlla90}a and \ref{fig:dlla34}a. In the meson sector,
we find that $\lambda$ is constrained to lie between
0.04--0.12, with $\lambda = 0.08$ giving the best fit to the data.
The overall shape of meson data is very well
realised at either energy, as can be seen from Figs. \ref{fig:dlla90m}b
and \ref{fig:dlla34m}b. The model parameters yield a reasonable fit to
all baryon and meson data at these two different energies. The
meson data is better fitted than the baryon data at both
the energies. The discrepancy in overall normalisation could be due to
the inclusive nature of the measurement (especially acute in the case of
$p$ and $\Lambda$) and possible energy dependence of the 
suppression factor $\lambda$. However, we emphasise that our model is 
fairly simple; its biggest advantage is that it predicts the production
rates of several mesons and baryons with relatively few inputs. 

Recently, the total inclusive charged hadron cross section has been
measured at LEP at 161 GeV \cite{Data161}. We know from the multiplicity
data at the $Z^0$ pole that 81\% (91\%) of the charged particle
inclusive cross section is from pions (pions plus kaons). Specifically,
the total charged particle multiplicity at $\sqrt s = 91.2$ GeV is $21.4
\pm 0.02 \pm 0.43$ \cite{Mult}, of which $17.05 \pm 0.43$ are $\pi^{\pm}$
and $2.26 \pm 0.01 \pm 0.16 \pm 0.09$ are $K^{\pm}$ mesons. We therefore
compare the charged particle spectrum at 161 GeV (with multiplicity
$24.46 \pm 0.45 \pm 0.44$) with our predictions for $\pi^{\pm}$ and
$(\pi^{\pm} + K^{\pm})$; we expect the latter should saturate the data
to within 10\%. Our model shows excellent agreement with data, as can be
seen from Fig. \ref{fig:dlla161}.

We remark that the charge factors $c_q$/$\sum_q {c_q}$ for quarks $q
= u,d,s$, are very different for pure $Z^0$ and photon exchange. 
For instance, the charge factor for an $s$ quark is 1/6 at 34 GeV
and 13/36 at 91.2 GeV, more than a factor of two larger. On the other
hand, that for the $u$ quark is almost a factor of two smaller. This
means that the photon exchange data is more sensitive to $u$ quark
fragmentation than the $Z^0$ data. That the model predictions for
strange hadrons such as $\Lambda$ and $K$ (where the $u$ contribution is
suppressed by a factor of $\lambda$) are systematically smaller
than data may therefore mean that $D_u$ is actually larger than the
model prediction, thus indicating that a single strangeness suppression
factor $\lambda$ may not suffice. In other words, our simple model may not
completely account for all SU(3) breaking effects. In this context,
it would be interesting to obtain data on the $\Sigma$ baryon at a
different energy and check whether this trend is visible there as well. 


Finally, the data (specially for mesons, $p$ and $\Lambda$) show a
decreasing trend at low $x$. The usual DGLAP evolution \cite{Dglap}
cannot account for such a trend since the pole in the splitting function
$P_{gg}$ always drives the gluon, and hence the sea, to larger values
at small $x$. In 1988, Dokshitzer, Khoze, and Troyan \cite{Dok}  proposed
a model wherein this dip could be
accounted for by including gluon coherence effects. The resulting
modified leading log approximation (MLLA) then gives a gaussian
distribution for the singlet fragmentation functions. In the next section,
we discuss singlet evolution using MLLA and look for improved fits to the
low $x$ data.

\subsection{Modified Leading log evolution}

The main result of the MLLA evolution \cite{Dok,Bassetto} is that the
low-$x$ singlet
fragmentation functions have a gaussian form in the variable $\log(x_p)$:
\begin{eqnarray}
x_pD(x_p,Q)= {N(Q)\over \sqrt{2\pi}\sigma(Q)} \exp\left[-\left[\log(x_p)
		-\log(x_0)\right]^2/[2\sigma^2(Q)]\,\right]~,
\end{eqnarray}
where $N(Q)$ is the total multiplicity, $\sigma$ is the width of the
gaussian and
${x_0}$ is the position of the peak of the gaussian. The $Q^2$
dependence of $N$, $\sigma$ and $x_0$ are computable for total inclusive
hadrons within this approach. They are given as an expansion in terms of
the scale parameter, $Y = \log(Q/\Lambda)$, $\Lambda = 200$ MeV: 
\begin{eqnarray} \nonumber
N(Q) & \propto & Y^{-B/2+1/4}~\exp \sqrt{16 N_c Y/b}~; \\ 
\sigma^2 & = & Y^2/(3z)~; \\ \nonumber 
\log(1/x_0) & = & Y\left[1/2+\sqrt{c/Y}-c/Y \right]~, \nonumber 
\end{eqnarray} 
where $N_c$ and $n_f$ are the number of colours and flavours, which
determine the constants,
$$
\begin{array}{rclrcl}
a & = & 11 N_c/3 + 2 n_f/(3 N_c^2)~; & b & = & (11 N_c - 2 n_f)/3~; \\
B & = & a/b~;                        & z & = & \sqrt{(16 N_c Y)/b}~; \\
c & = & (11/48) \, \left[1+(2 n_f)/(11 N_c^3)\right]^2 & / &
               \multicolumn{2}{l}{\left[1-2 n_f/(11 N_c) \right]~.}
\end{array}
$$
The total multiplicities (at 91.2 GeV, for instance)
\cite{Data90,Sigdata,Mult} can be used to fix the proportionality constant
for $N(Q)$; the individual particle multiplicities then determine
$N^h(q)$, the multiplicity of the specific hadron, $h$. The values of
$\sigma$ and $x_0$ are in good agreement with inclusive data \cite{Dok};
however, we are here interested in semi-inclusive spectra. In general,
the peak shifts to smaller $x$ (here meaning $x_p$) values for heavier
hadrons.  Also, the semi-inclusive widths are naturally smaller than
the total inclusive ones. We therefore parametrise the corresponding
semi-inclusive parameters as
\begin{eqnarray} \nonumber
x_0^h & = & C_1^h x_0~, \\ \nonumber
\sigma^2_h & = & C_2^h \sigma^2~, \nonumber
\end{eqnarray}
where $x_0^h$ is the peak position and $\sigma_h$ the width of the data
for hadron $h$. Here $C_1^h$ and $C_2^h$ are $Q^2$ independent constants
which we fit to the 91.2 GeV data. They are given in Table
\ref{tab:inputml}. These are
then used to determine the rates at lower energies. The resulting fits
are again quite good and shown in Figs. \ref{fig:mlla90} and
\ref{fig:mlla34} for 90 and 34 GeV respectively. Note that the cross
section has been plotted as a function of $x_p$ and not $x_E$ here.

Note that the MLLA is a fit to the singlet fragmentation functions alone;
therefore comparison should be made with data for
$x_p$ \raisebox{-0.4ex}{$\stackrel{<}{\scriptsize \sim}$} 0.1, where the
valence contribution is expected to be small. In 
the case of $Z^0$ exchange, this is also a good fit to the entire
data. This is because at 91.2 GeV, the cross section is dominated by the
singlet term, as can be seen from writing Eq. \ref{eq:sns} explicitly:
$$
\by{1}{\sigma_{tot}} \by{\d \sigma^{h,Z}}{\d x} = \by{12
\Sigma^h - 1.5 D_3 - 0.5 D_8}{36}~.
$$
We see that the singlet contribution is about 10 times larger than
either of the octet contributions. We therefore expect the MLLA approach
to yield sensible fits to the data at this energy. In the case of photon
exchange data, the singlet contribution is still large:
$$
\by{1}{\sigma_{tot}} \by{\d \sigma^{h,\gamma}}{\d x} = \by{2
\Sigma^h + 1.5 D_3 + 0.5 D_8}{6}~,
$$
but the $D_3$ contribution is not small. Hence MLLA may not be a very
good description of the data at smaller energies, especially at larger
$x$. However, for the case
of $\Lambda$ and $\Sigma^0$, $D_3 = 0$ so that the MLLA singlet term
still saturates the event rate to a good approximation. 

\section {Discussion and Conclusions}

We have proposed a simple model for quark fragmentation into an octet
baryon or a pseudoscalar meson, using SU(3) symmetry of quarks and octet
hadrons. All quark fragmentation functions have been described in terms
of three SU(3) symmetric functions $\al(x,Q^2)$, $\be(x,Q^2)$ and
$\ga(x,Q^2)$ and an SU(3) breaking parameter $\lambda$. The antiquark 
fragmentation functions are correspondingly described by $\overline{\al}$,
$\overline{\be}$ and $\overline{\ga}$. There are 3 quark (plus 3 antiquark) 
fragmentation functions corresponding to a given hadron; hence a given
hadron octet would involve a total of $(24 \times 2)$ fragmentation
functions. All these are described in our model by just 6 functions, leading
to a very simple model, but with strong predictive power. Leading log
evolution of these fragmentation functions has been used to compare the
model predictions with data. We find that it is possible to fit the
model parameters in such a way as to get a good agreement with the
$x$-dependence of all octet baryons and mesons, at two different sample
energies (corresponding to $Z^0$ and photon exchange) over
most of the $x$ range of available data. These fits were then used
to determine the inclusive cross section at 161 GeV where both photon
and $Z^0$ exchange are involved. There was good agreement with data here
as well. We have used both DGLAP evolution
as well as the modified leading log approximation (MLLA) to evolve the
fragmentation functions; the latter has especially been used to explain 
the decrease in the hadroproduction rates at small-$x$ seen in the data
for many hadrons. We have used a small non-zero input gluon distribution
(as shown in Table \ref{tab:inputll}); however, very little sensitivity to
the gluon fragmentation function is seen and this is therefore not
well-determined in our analysis. 

The model realises the shape of the $x$-distribution of all available data 
on octet mesons and baryons very well. It does not describe the $\Lambda$
and $K$ data at 34 GeV very accurately; however, it is able to
give a good agreement with even this data to within 2$\sigma$. All other
baryon and meson data are fitted very well.
However, we note that it is possible to get good fits at
all $Q^2$ for each hadron {\it individually}. The SU(3) symmetry constraint
relating the different hadron fragmentation functions worsens the fit
in some cases; this reflects the simplicity of our model, which
incorporates SU(3) symmetry breaking effects in a very simple way. The
goodness-of-fit from the model therefore also indicates the extent to
which this symmetry breaking is a universal phenomenon, independent of
the type of quark or diquark that is produced
\cite{Jetset,Herwig,Sigdata}. 

The parameter 
$\lambda$ takes into account the difference in masses of the strange and
non-strange quarks and suppresses non-strange quark fragmentation into
strange hadrons. This parameter is similar to the suppression factor of
the Lund Monte Carlo \cite{Lund}, however, it is determined by means of
a simple comparison of data of stange and non-strange hadrons. The Lund
model uses string fragmentation and has a much larger suppression for
the case of baryons (suppression factor = 0.06) as compared to the
suppression factor of 0.2-0.3 for mesons. Our model has very similar
values for the suppression factors for the two cases ($\lambda$ = 0.07 
for baryons, 0.08 for mesons). 

Another approach \cite{Nzar} uses an SU(6) analysis of fragmentation 
functions using a quark and diquark model. Our SU(3) symmetric functions 
$\al$ and $\be$ are analogous to the SU(6) symmetric functions
$\widehat{S}(z)$ and $\widehat{T}(z)$ defined therein. The function $\ga$
(not included in their model) describes sea fragmentation, for instance,
$s$ fragmenting to a proton. We find that $\ga$ is large in the small $x$
region, so that its contribution is significant and cannot be ignored.

We find that the strange quark fragmentation dominates strange
hadroproduction over almost the entire $x$ range. This is especially
true for $\Lambda$, which has recently been of much interest
\cite{Vogelsang,Ravi}. It is possible to extend the model to include
spin-dependent fragmentation functions; the unpolarised result then
indicates that polarised $\Lambda$ fragmentation will be dominated by
its strange fragmentation function, which can then be readily
parametrised and studied. 

Finally, our results suggest that there is indeed an underlying symmetry
among the baryons and mesons in an octet, which
can be tested further by extending the model to decuplet baryons and
other hadrons.

\vspace{0.5cm}

\paragraph{Acknowledgments} : 
We are grateful to the late K.V.L. Sarma for helpful discussions.
We would like to thank B.D. Roy and Debajyoti Choudhury for active 
discussions and ideas. A.R. thanks
U.G.C. for grant of a fellowship, R.K. Shivpuri for constant support 
and encouragement and MRI for hospitality.

\vspace{1cm}

\noindent {\Large \bf Appendix A}

\vspace{0.3cm}

\noindent We briefly detail the calculations leading to the results in
Table \ref{tab:frag} for the quark fragmentation functions in terms
of $\alpha$, $\beta$ and $\gamma$. 

Let $q_i$ be a quark triplet, $B_i^j$ the hadron octet, $i,j,k = 1,2,3$. 

\paragraph{Case 1 $X$ is a triplet, $X_i$:}
Then the invariant amplitude for the
process $q~\rightarrow~H+X$ is $q_i H_j^i X^j$, where $X_i$ and $q_i$
are normalised. Here $H_i^j$ are the elements of the meson/baryon matrix
(see Table \ref{tab:octet}). Thus, the rate for $u~\rightarrow~p+X$
is $\alpha \vert uB_3^1X^3 \vert^2$ which is equal to $\alpha$. Similarly,
the rate for $u~\rightarrow~\Lambda+X$ is $\alpha\vert uB_1^1X^1 \vert^2$
which is equal to $\alpha$/6 and so on. 
  
\paragraph{Case 2 $X$ is a sixplet, $X_{ij}$:}
Now $X_{ij}$ is symmetric in $i$ and $j$ and is expressed in terms of
triplets as $(q_iq_j + q_jq_i)/\sqrt 2$, where each $q$ is
normalised. The invariant amplitude is $\epsilon^{imj} q_i H_j^k
X_{km}$. Thus, $d~\rightarrow~p + X$ as $\beta\vert {\sqrt 2} \vert^2$
and so on.

\paragraph{Case 3 $X$ is a fifteenplet, $X_i^{jk}$:}
Then $X_i^{jk}$ is symmetric in $j, k$ and is antisymmetric in $i, j$.
In terms of triplets, the normalised $X$ can be re-expressed as 
$$
X_i^{jk}  =  {1 \over \sqrt 2}\left[q^jq^kq_i + q^kq^jq_i - {1 \over 4} 
\delta_i^j (q^lq^kq_l+q^kq^lq_l) - {1 \over 4}
\delta_i^k (q^j q^l q_l + q^l q^j q_l)\right]~,
$$
where each of the $q_i$'s is normalised. 

The invariant amplitude is $X_j^{ik}H_i^jq_k$. In this case, we shall
have to take the interference terms for the diagonal elements of the 
meson/baryon matrix also into account. Note that $X_i^{ij}$ = 0 (sum over 
$i$ is implied). Thus, the rate for $u~\rightarrow~\Lambda + X$ is
$\gamma\vert (1 /\sqrt 6) X_1^{11} +(1 / \sqrt 6) X_2^{21} -(2 / \sqrt
6) X_3^{31} \vert^2$ which is equal to
$\gamma\vert (3 / \sqrt 6)(X_1^{11}+X_2^{21}) \vert^2$. On evaluating
this expression, we find that this is equal to 9/8$\gamma$. The
other rates can also be found in a similar manner. The final results
are given in Table \ref{tab:frag}.

\newpage

\begin{table}[htb]
\centering
$$
\begin{array}{ccc}
\left( \begin{array}{ccc} 
\by{\Sigma^0}{\sqrt{2}}+\by{\Lambda}{\sqrt{6}} & \Sigma^+ & p \\
\Sigma^- & \by{-\Sigma^0 }{ \sqrt{2}}+\by{\Lambda }{ \sqrt{6}} & n \\
\Xi^- & \Xi^0 & \by{-2 \Lambda }{ \sqrt{6}}
\end{array} \right) & \hspace{1.2cm} & 
\left( \begin{array}{ccc} 
\by{\pi^0}{\sqrt{2}}+\by{\eta}{\sqrt{6}} & \pi^+ & K^+ \\
\pi^- & \by{-\Sigma^0 }{ \sqrt{2}}+\by{\eta }{ \sqrt{6}} & K^0 \\
K^- & \bar{K^0} & \by{-2 \eta }{ \sqrt{6}} 
\end{array} \right)
\end{array}
$$
\caption{(a) Members of the meson octet and (b) Members of
the baryon octet.}
\label{tab:octet}
\end{table}

\vspace{1.5cm}

\begin{table}[htb]
\centering
\begin{tabular}{|ccl|ccl|} \hline
fragmenting & \multicolumn{2}{c|}{${}_{\displaystyle p/K^+}$} & fragmenting &
\multicolumn{2}{c|}{${}_{\displaystyle n/K^0}$} \\
quark & & & quark & & \\  \hline
$u$ & : &  ${\alpha}+{\beta}+{\frac{3}{4}}{\gamma}$ & 
$u$ & : &  $2{\beta}+{\gamma}$ \\
$d$ & : &  $2{\beta}+{\gamma}$ & 
$d$ & : &  ${\alpha}+{\beta}+{\frac{3}{4}}{\gamma}$ \\
$s$ & : &  $2 {\gamma}$ & 
$s$ & : &  $2 {\gamma}$ \\ \hline
fragmenting & \multicolumn{2}{c|}{${}_{\displaystyle \Lambda^0/\eta}$} & fragmenting &
\multicolumn{2}{c|}{${}_{\displaystyle \Sigma^0/\pi^0}$} \\
quark & & & quark & & \\  \hline
$u$ & : &  
$\frac{1}{6}{\alpha}+\frac{9}{6}{\beta}+\frac{9}{8}{\gamma}$ &
$u$ & : &  
$\frac{1}{2}{\alpha}+\frac{1}{2}{\beta}+\frac{11}{8}{\gamma}$ \\
$d$ & : &  
$\frac{1}{6}{\alpha}+\frac{9}{6}{\beta}+\frac{9}{8}{\gamma}$ &
$d$ & : &  
$\frac{1}{2}{\alpha}+\frac{1}{2}{\beta}+\frac{11}{8}{\gamma}$ \\
$s$ & : &  
$\frac{4}{6}{\alpha}+\frac{9}{6}{\gamma}$ &
$s$ & : &  $2\beta+\gamma$ \\ \hline
fragmenting & \multicolumn{2}{c|}{${{}_{\displaystyle \Sigma^+/\pi^+}}$} &
fragmenting &  \multicolumn{2}{c|}{${{}_{\displaystyle \Sigma^-/\pi^-}}$} \\
quark & & & quark & & \\  \hline
$u$ &  : & ${\alpha}+{\beta}+{\frac{3}{4}}{\gamma}$ & 
$u$ &  : & $2 {\gamma}$ \\ 
$d$ &  : & $2 {\gamma}$ & 
$d$ &  : & ${\alpha}+{\beta}+{\frac{3}{4}}{\gamma}$ \\ 
$s$ &  : & $2{\beta}+{\gamma}$ & 
$s$ &  : & $2{\beta}+{\gamma}$ \\ \hline
fragmenting & \multicolumn{2}{c|}{${{}_{\displaystyle \Xi^0/\overline{K^0}}}$} &
fragmenting & \multicolumn{2}{c|}{${{}_{\displaystyle \Xi^-/K^-}}$} \\
quark & & & quark & & \\  \hline
$u$ & : & $2{\beta}+{\gamma}$ & 
$u$ & : & $2 {\gamma}$ \\ 
$d$ & : & $2 {\gamma}$ & 
$d$ & : & $2{\beta}+{\gamma}$ \\ 
$s$ & : & ${\alpha}+{\beta}+{\frac{3}{4}}{\gamma}$ & 
$s$ & : & ${\alpha}+{\beta}+{\frac{3}{4}}{\gamma}$ \\
\hline

\end{tabular}
\caption{Quark fragmentation functions into members of the
baryon and meson octet in terms of the SU(3) functions, $\alpha$,
$\beta$ and $\gamma$. }
\label{tab:frag}
\end{table}

\vspace{1cm}

\begin{table}[htb]
\centering
\vspace{1cm}

\begin{tabular}{|c|c|c|c|c|} \hline
      & {$\alpha_V$} & {$\beta_V$} & {$\gamma$} & {$g$} \\ \hline
$a$ & 3.0 & 25.8 & 3.5 & 2.5 \\ \hline
$b$ & 4.8 & 8.0 & 13.6 & 13.4 \\ \hline
$c$ & -0.55 & 1.52 & 0.12 & 0.12 \\ \hline
$d$ & -3.96 & -5.19 & -7.82 & 0 \\ \hline
$e$ & 13.12 & 9.84 & 38.0 & 0 \\ \hline
\end{tabular}

\vspace{0.5cm}

\begin{tabular}{|c|c|c|c|} \hline
      & {$V$} & {$\gamma$} & {$g$} \\ \hline
$a$ & 2.33 & 3.5 & 0.25 \\ \hline
$b$ & 2.15 & 12.76 & 11.4 \\ \hline
$c$ & -0.64 &-0.75 & 0.12 \\ \hline
$d$ & 5.35 & 3.87 & 0 \\ \hline
$e$ & -5.12 & 61.59 & 0 \\ \hline
\end{tabular}

\caption{Input values at $Q^2 = 2$ GeV${}^2$ for the valence and singlet
fragmentation functions for the (a) meson and (b) baryon octet.}
\label{tab:inputll}

\vspace{3cm}

\end{table}

~
\vspace*{2cm}

\begin{table}[htb]
\centering
\begin{tabular}{|c|c|c|c|c|} \hline
      & {$p$} & {$\Lambda$} & {$\Sigma^{\pm}$} & {$\Xi^-$} \\ \hline
$C_1$ & 0.70 & 0.70 & 0.70 & 0.70 \\ \hline
$C_2$ & 0.4 & 0.4 & 0.4 & 0.4 \\ \hline
\end{tabular}
\begin{tabular}{|c|c|c|c|c|} \hline
 & {$\pi^{\pm}$} & {$\pi^0$} & {$K^{\pm}$} & {$K^0$} \\ \hline
$C_1$ & 0.92 & 0.92 & 0.75 & 0.75 \\ \hline
$C_2$ & 0.59 & 0.59 & 0.425 & 0.425 \\ \hline
\end{tabular}

\caption{Multiplying constant factors for description of singlet
fragmentation functions for mesons and baryons in the MLLA
approach.}
\label{tab:inputml}
\end{table}


\newpage

\newpage
\vspace{0.3cm}

\addtocounter{figure}{-1}
\begin{figure}[ht]
\centering
\vskip 15truecm

\includegraphics{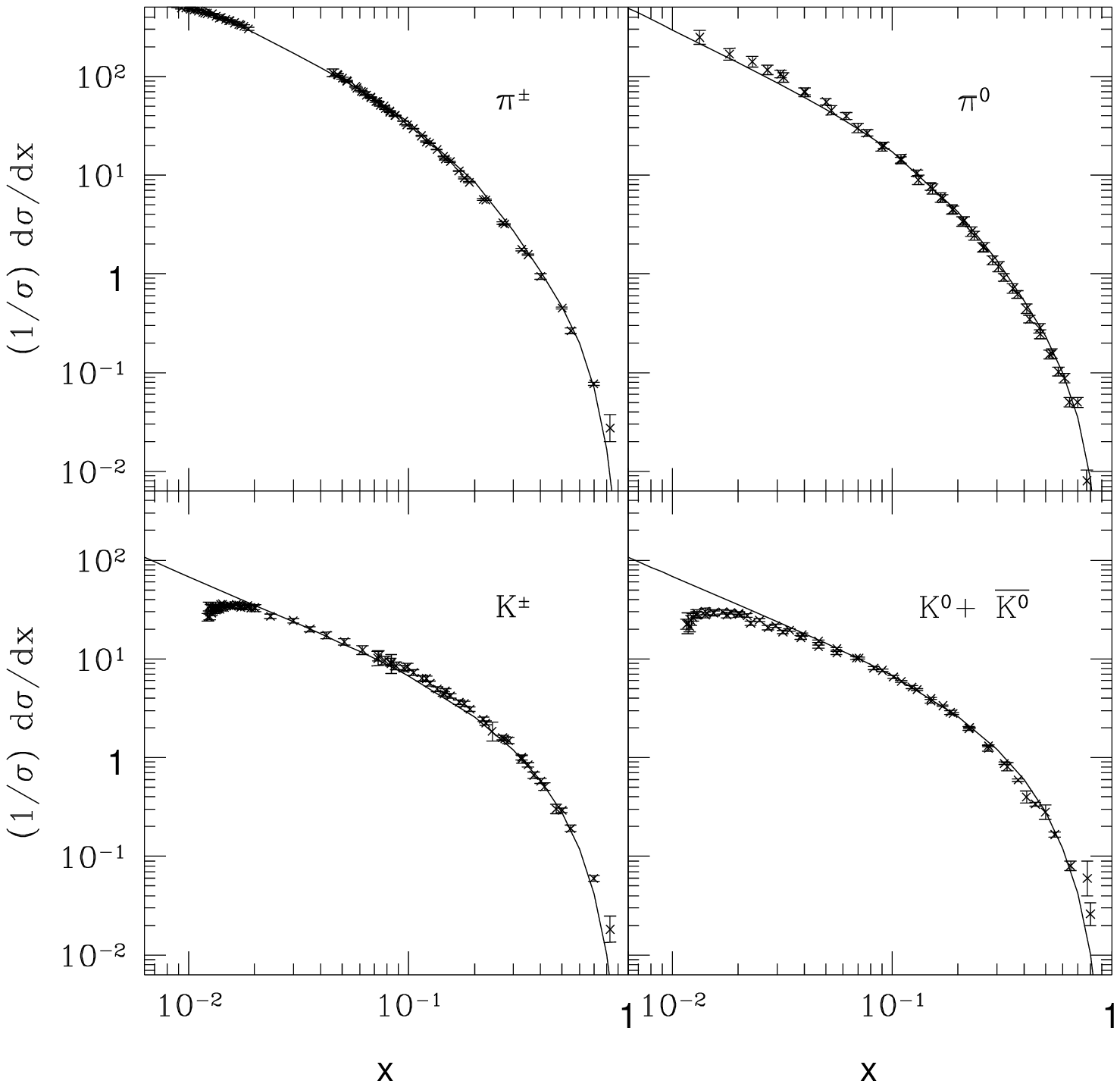}
\caption[dummy]{(b) The figure shows the model fits to the 90 GeV data
for mesons using DGLAP evolution.}

\label{fig:dlla90m}
\end{figure}

\newpage
~
\begin{figure}[ht]
\centering
\vskip 10truecm

\includegraphics{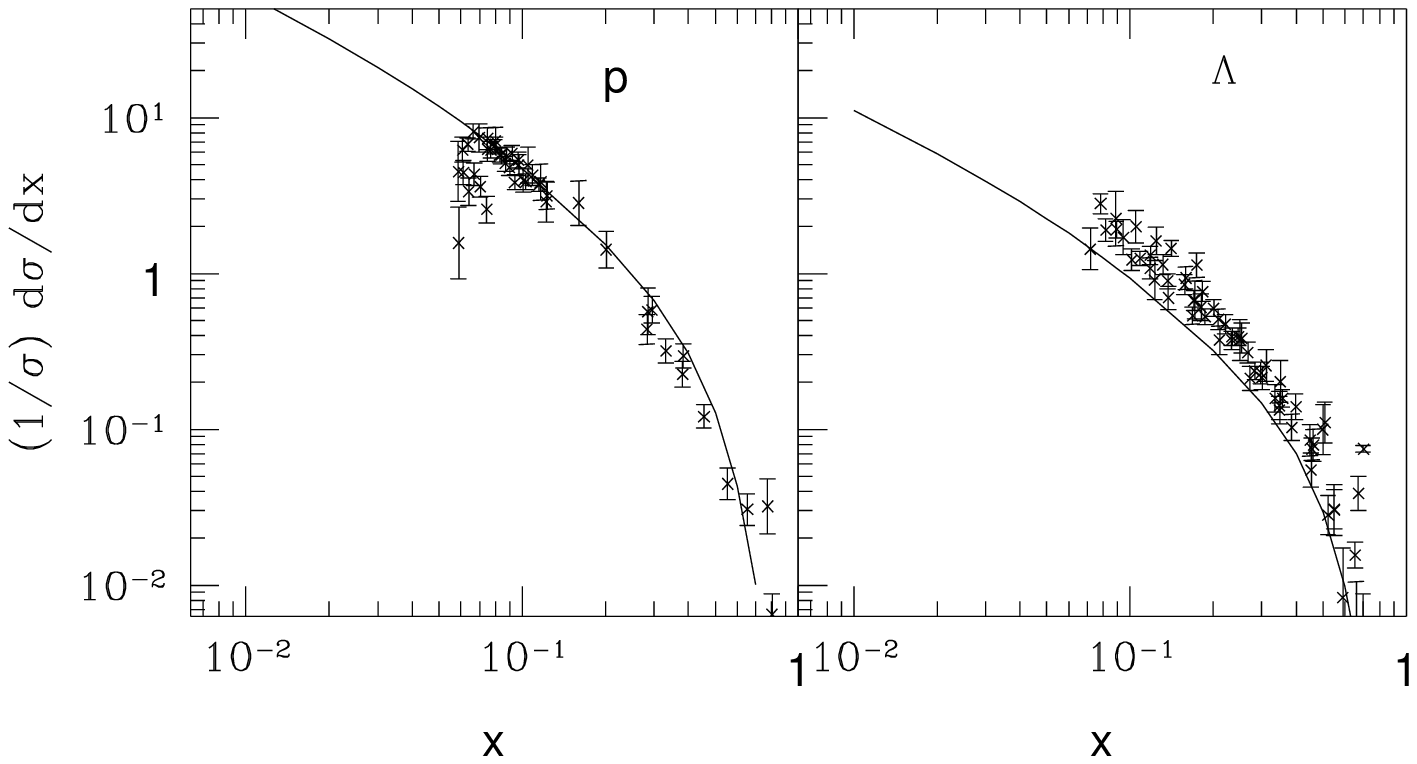}
\caption[dummy]{(a) The figure shows the model fits to the 34 GeV data
for baryons using DGLAP evolution.}

\label{fig:dlla34}
\end{figure}

\addtocounter{figure}{-1}
\begin{figure}[ht]
\centering
\vskip 10truecm

\includegraphics{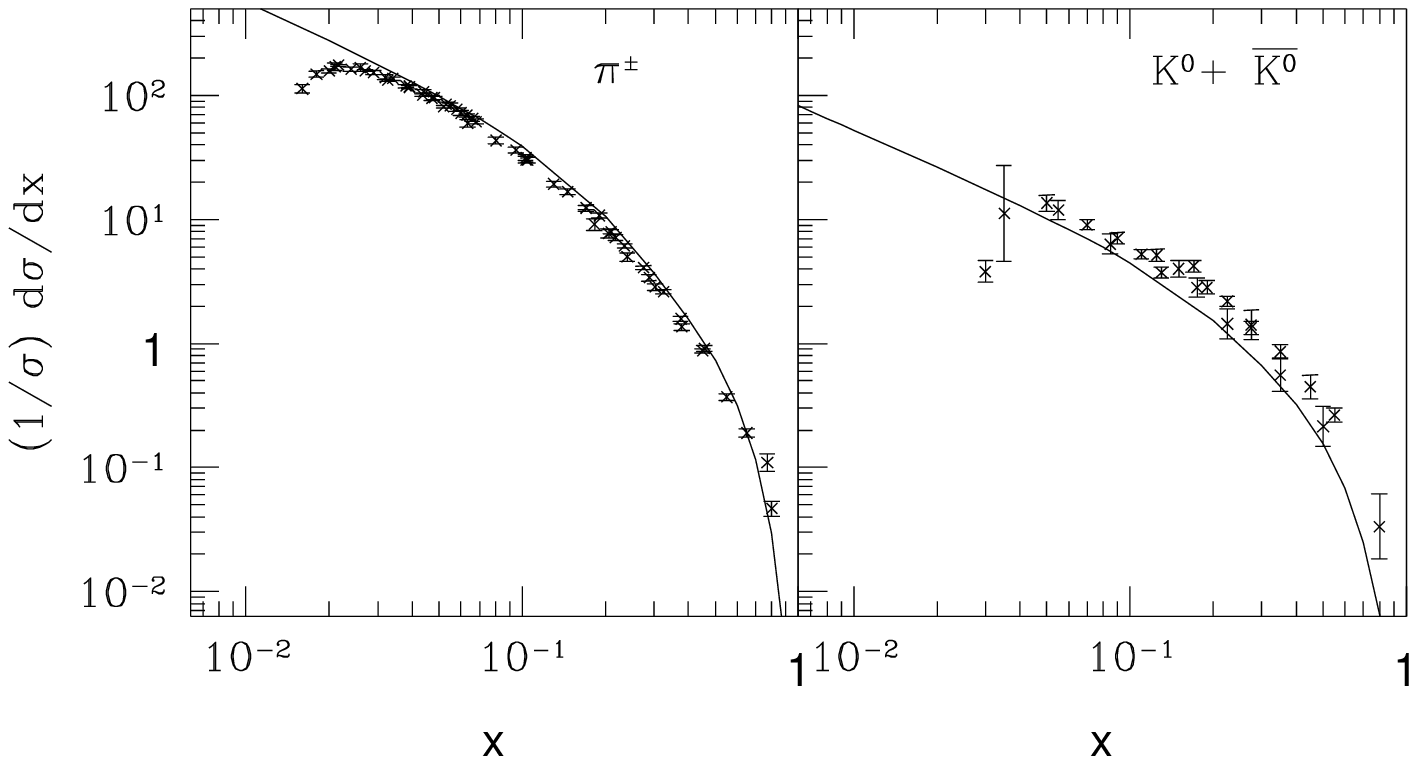}
\caption[dummy]{(b) The figure shows the model fits to the 34 GeV data
for mesons using DGLAP evolution.}

\label{fig:dlla34m}
\end{figure}

\newpage
\begin{figure}[ht]
\centering
\vskip 10truecm

\includegraphics{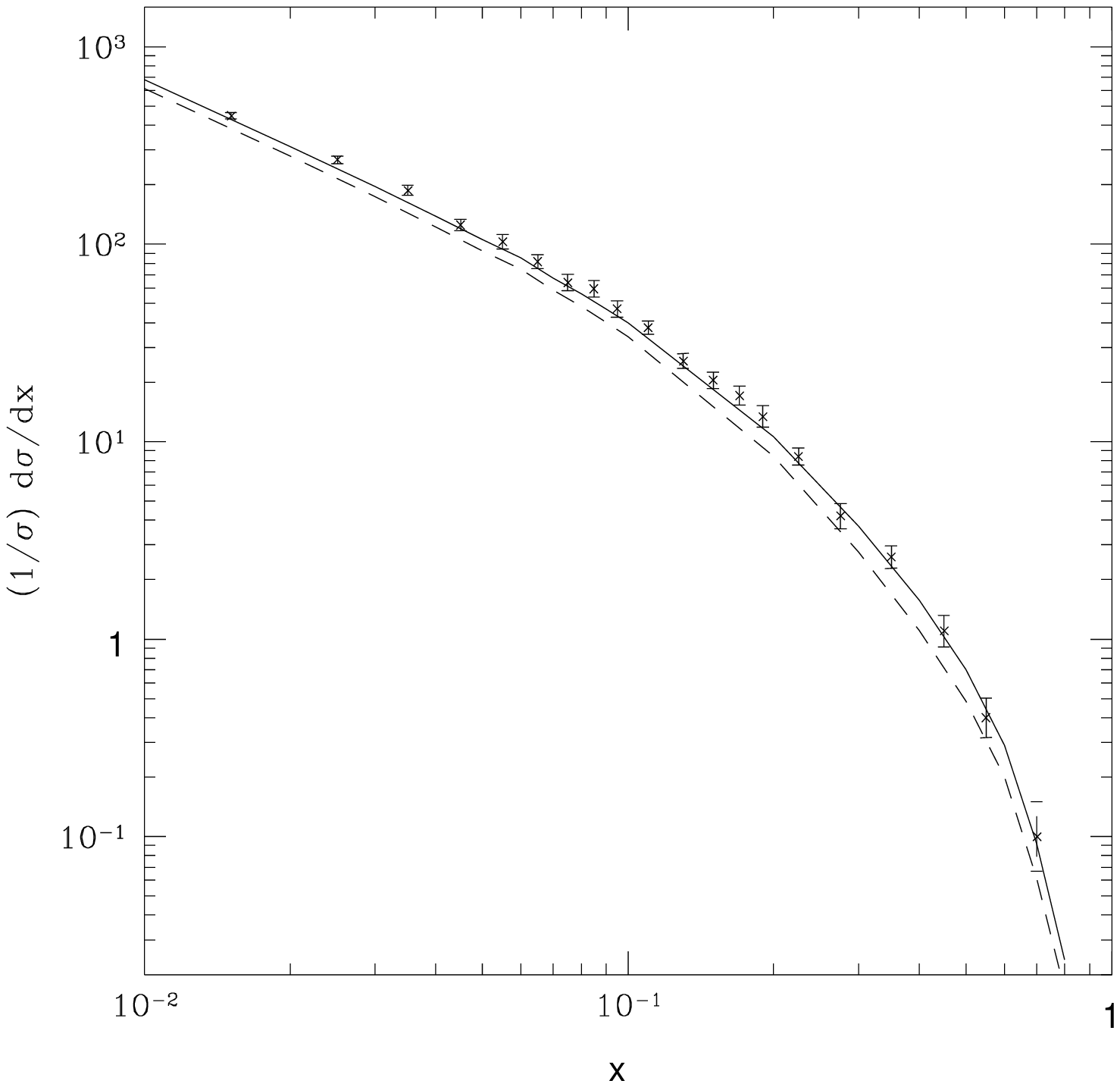}
\caption[dummy]{The model prediction for
$\pi^{\pm}$ (dotted lines) and $\pi^{\pm}+K^{\pm}$ (solid lines)
is compared with the total inclusive charged particle data at 161 
GeV \cite{Data161}.}

\label{fig:dlla161}
\end{figure}

\newpage
\vspace{0.3cm}
\begin{figure}[ht]

\centering
\vskip 15truecm

\includegraphics{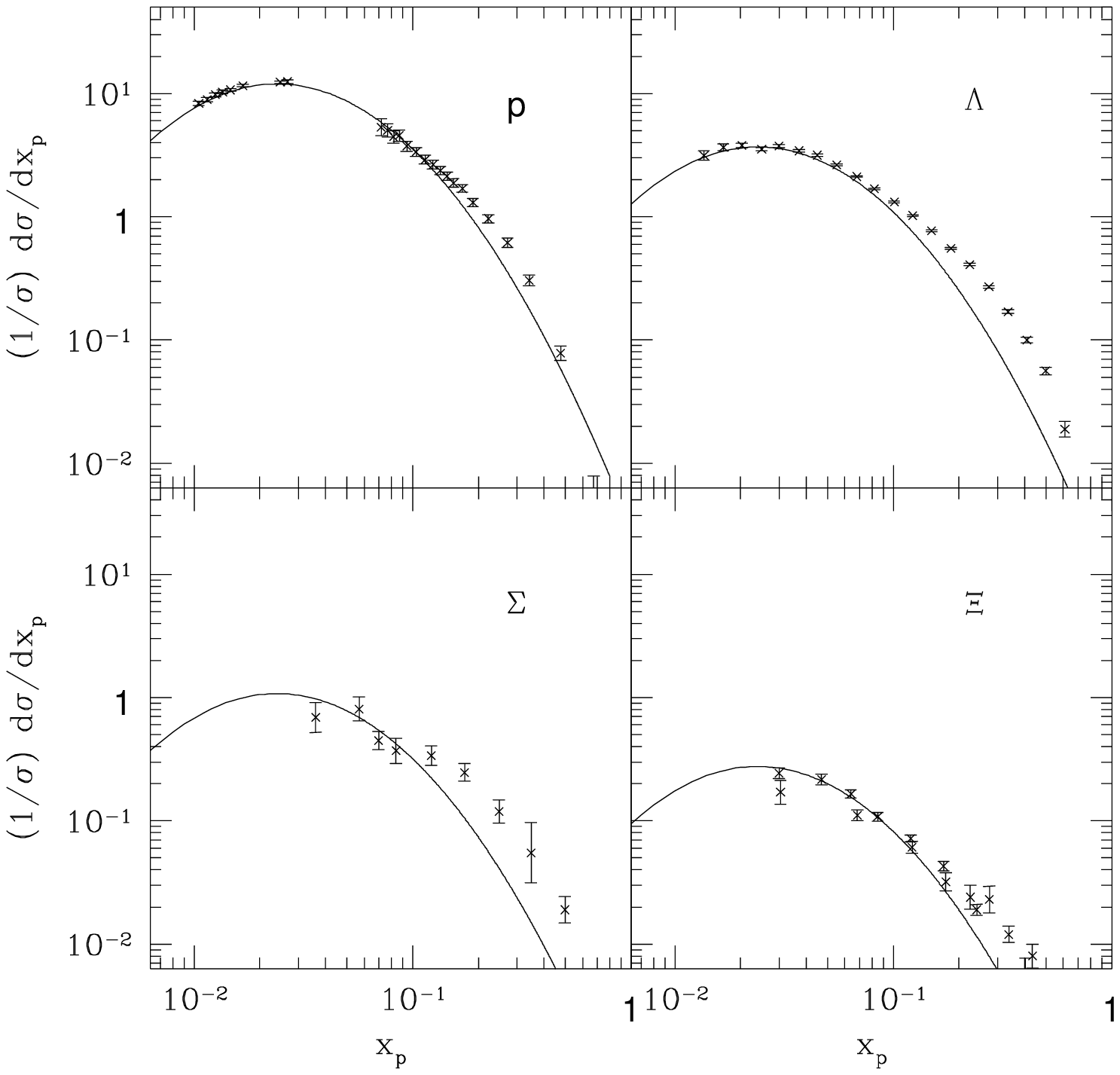}
\caption[dummy]{(a) The figure shows the model fits to the 90 GeV data
for baryons using MLLA evolution. Note that we have used $x = x_p$ in 
order to clearly exhibit the small-$x$ data which is of interest here.}

\label{fig:mlla90}
\end{figure}

\newpage

\addtocounter{figure}{-1}
\begin{figure}[ht]

\centering
\vskip 15truecm

\includegraphics{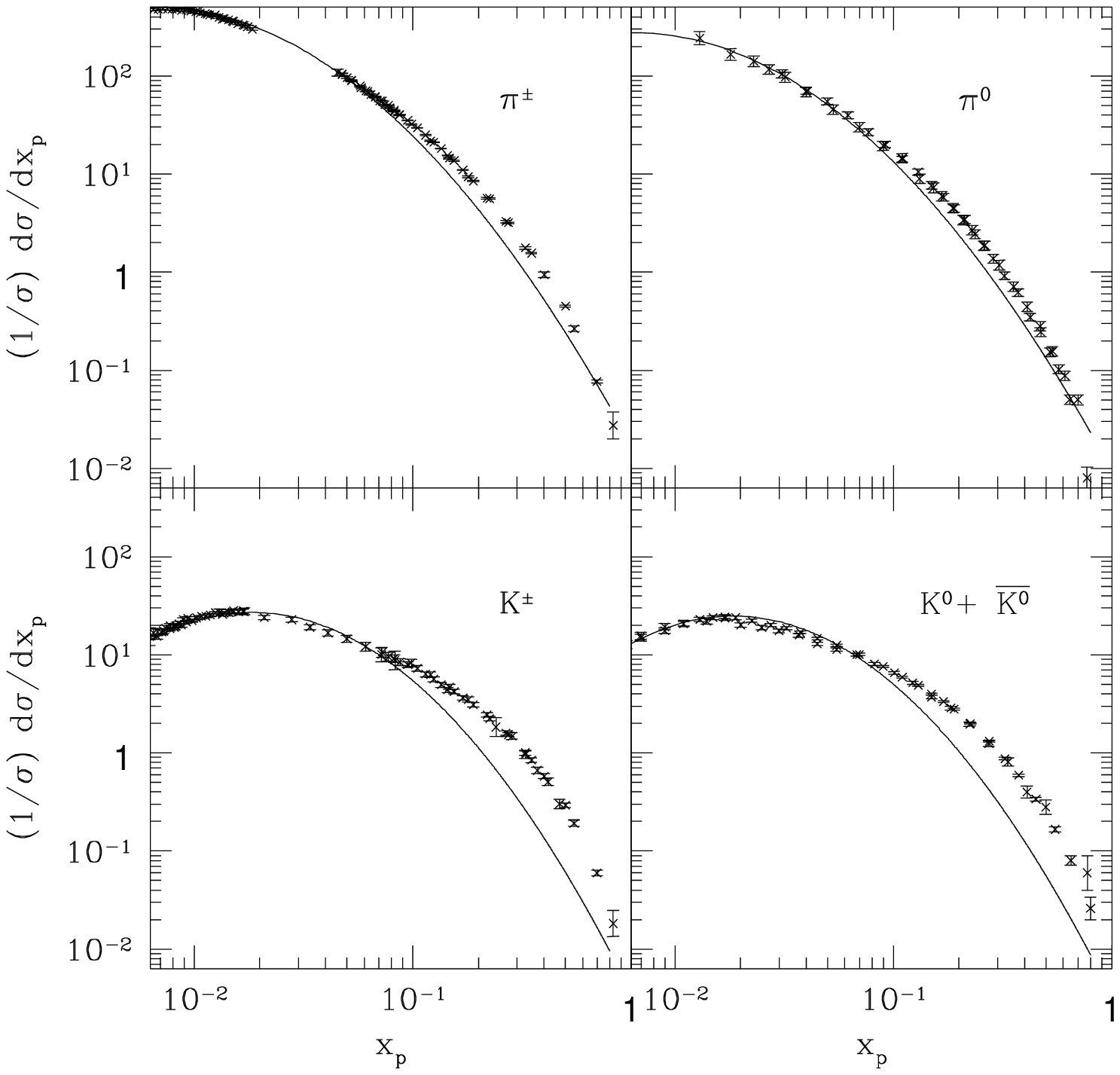}
\caption[dummy]{(b) The figure shows the model fits to the 90 GeV data
for mesons using MLLA evolution. Note that we have used $x = x_p$ in 
order to clearly exhibit the small-$x$ data which is of interest here.}

\label{fig:mlla90m}
\end{figure}

\begin{figure}[ht]
\centering

\vskip 9truecm

\includegraphics{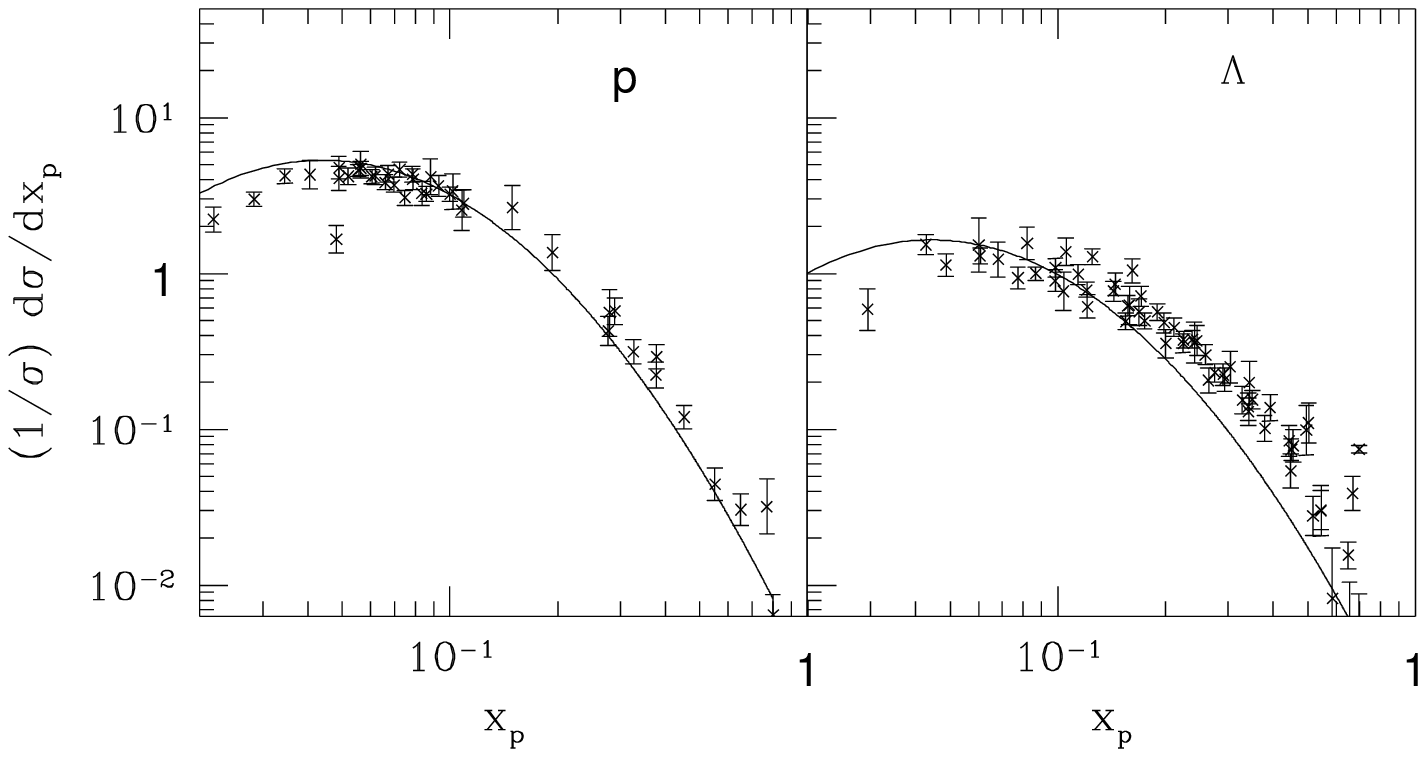}
\caption[dummy]{(a) The figure shows the model fits to the 34 GeV data
for baryons using MLLA evolution. Note that we have used $x = x_p$ in 
order to clearly exhibit the small-$x$ data which is of interest here.}

\label{fig:mlla34}
\end{figure}

\addtocounter{figure}{-1}

\begin{figure}[htb]
\centering

\vskip 9truecm

\includegraphics{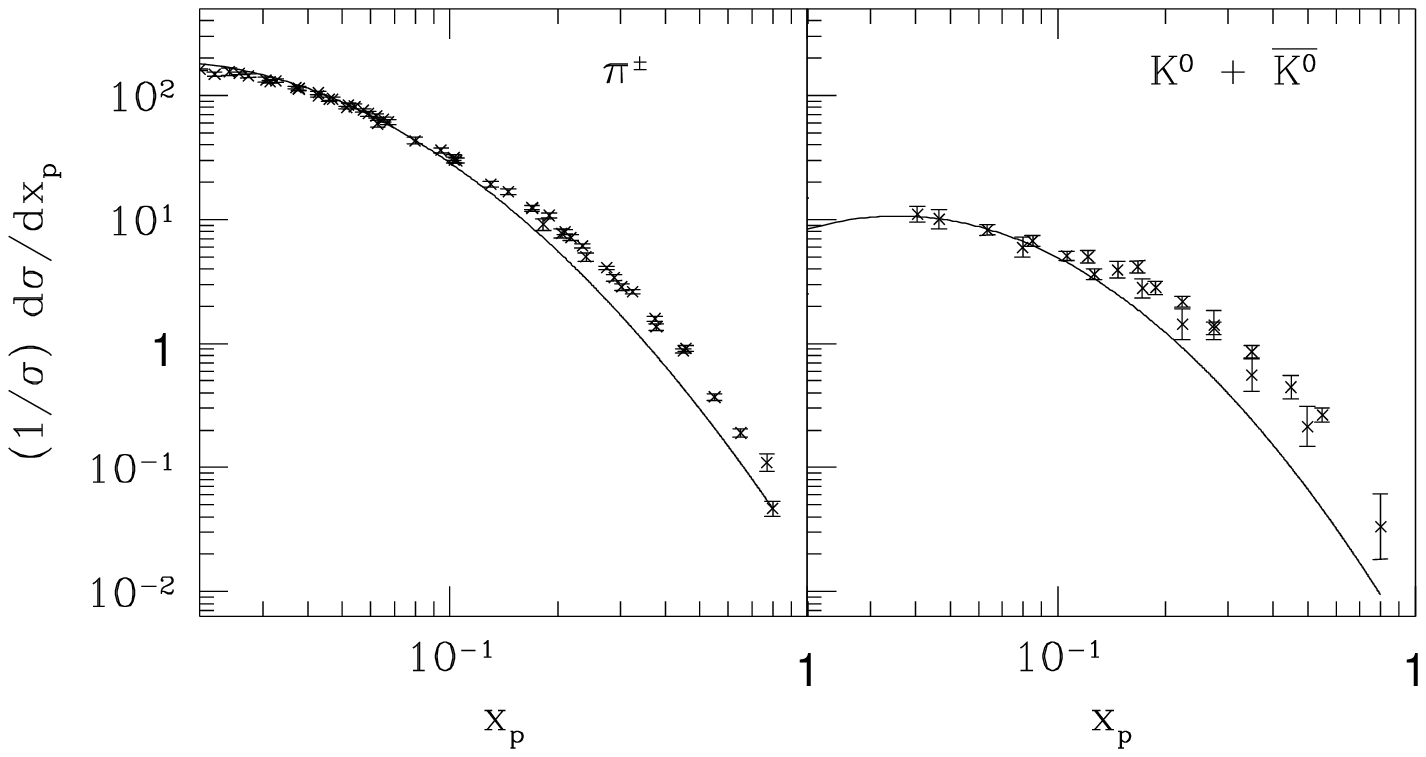}
\caption[dummy]{(b) The figure shows the model fits to the 34 GeV data
for mesons using MLLA evolution. Note that we have used $x = x_p$ in 
order to clearly exhibit the small-$x$ data which is of interest here.}

\label{fig:mlla34m}
\end{figure}

\end{document}